\documentstyle[12pt]{article}
\begin{document}

\setlength{\textwidth}{15truecm}
\setlength{\textheight}{23cm}
\baselineskip=24pt

\centerline{\bf {Spiral-like light profiles but elliptical-like kinematics
in mergers of galaxies}}

\bigskip

\bigskip

\centerline {\bf{Chanda J. Jog and Aparna Chitre}}

\medskip

\centerline {Department of Physics, Indian Institute of Science,
Bangalore 560012, India}
\centerline {{e-mail: cjjog@physics.iisc.ernet.in,aparna@physics.iisc.ernet.in}}

\bigskip

\centerline{ To Appear in A \& A Letters, 393, Oct. 2002}

\newpage

\centerline {\bf Abstract}

It is commonly accepted that a merger of two spiral galaxies  results 
in a remnant  with an elliptical-like surface-brightness 
profile. Surprisingly, our recent study (Chitre \& Jog 2002) of the 2MASS
data for twenty-seven advanced mergers of galaxies has shown that 
half of these have a light distribution that decreases exponentially with 
radius. Such a 
distribution normally characterizes a rotationally supported disk in a 
spiral galaxy. Here we show from kinematic data for two of these mergers, 
Arp 224 and Arp 214, that the main support 
against gravitational collapse comes from pressure due to random motion
of stars as seen in an elliptical galaxy rather than from rotation.
The origin of the unusual combination of properties seen here is
a puzzle. The  standard theoretical
N-body models in the literature cannot account for these systems.
 Further observational and dynamical studies of this new class of 
merger remnants are needed, and would be important for understanding 
merger dynamics and galaxy evolution.

\noindent Key words. {galaxies: elliptical - galaxies: evolution - galaxies:
 interactions -  galaxies: kinematics and dynamics - galaxies: spiral  - 
galaxies: structure}

\noindent Running Title: {Spiral-like light profiles but elliptical-like
kinematics in mergers}

\noindent {\bf 1. Introduction}

It is well-established observationally that a merger of two gas-rich
spirals can result in  an $r^{1/4}$
de Vaucouleurs radial profile typical of ellipticals (Schweizer
1982, Wright et al. 1990, Stanford \& Bushouse 1991). Our recent study 
(Chitre \& Jog 2002) has shown the
existence of robust, exponential profiles in surface brightness
in thirteen advanced mergers of spiral galaxies, similar to the radial
profiles seen in isolated spiral galaxies (Freeman 1970). This result is
unexpected and is in contrast to the result from the
earlier studies of mergers (Wright et al. 1990, 
Scoville et al. 2000, Genzel et al. 2001). We note that these
earlier studies had mainly concentrated on infrared-bright galaxies
whereas our sample was selected purely based on a disturbed
appearance without any bias towards IR luminosity.
 The  existence of exponential profiles, that too
in a large fraction of the sample, is a puzzle and needs further
dynamical study.

To get a coherent picture of the mergers with exponential profiles, 
in this Letter, we study the complementary information from kinematics 
available for two of these 
galaxies, namely Arp 224 (NGC 3921) and Arp 214 (NGC 3718).
We also obtain additional photometric properties of the thirteen
galaxies with exponential profiles by analyzing the 
$K_{\mathrm s}$ band images from the Two Micron All Sky Survey (2MASS).
 The most significant result from this paper is that
these merger remnants appear to be largely pressure-supported as
seen in an elliptical galaxy and yet show an exponential mass profile as
seen in a spiral galaxy. 

Section 2 gives the kinematic data for the above two galaxies and the results 
deduced, and also the photometric properties of
the entire sample.
Section 3 gives the dynamical implications and puzzles from these
results, and Sect. 4 contains a summary and discussion.

\bigskip

\noindent {\bf 2. Results on Photometric and Kinematic Properties} 

The kinematical data on rotation and random motion for Arp 224 
and Arp 214 are taken from the HYPERCAT
archival database (Simien \& Prugniel 1997, Heraudeau \& Simien 1998).
These observations involve absorption spectroscopic
measurements and a Fourier-fitting analysis and give the
rotation velocity projected along the major axis and the
velocity dispersion around it with an accuracy of dispersion of $<$ 30 
km s$^{-1}$.

Figures 1 and 2 show  the variation in the surface
brightness ($\mu_{\mathrm K}$), the circular rotation velocity
($V_{\mathrm c}$), and the 
velocity dispersion ($\sigma$), with radius 
for Arp 224 and Arp 214 respectively. The surface brightness
profile was derived from the 2MASS data
(Figs. 1a and 2a) and it shows a robust exponential fit (see Chitre \& Jog 
2002 for details). On doing a
similar fitting for the J and the H band data from 2MASS, we get
disk scalelengths that are comparable within the error bars (of $\leq$
10 \%), with no clear dependence on wavelength. Thus, the
luminosity profile is not affected significantly by the dust
extinction within the two galaxies.

The most striking result from Figs. 1c and 2c is that the 
velocity dispersion values are
large $\sim$ 100 km s$^{-1}$ within a few kpc of the galaxy centre.
Thus, these merger remnants are largely pressure-dominated as seen in
an elliptical galaxy and yet surprisingly show an exponential mass profile as
seen in an isolated spiral galaxy. Given the high velocity dispersion
values, the mass distribution is probably not
confined to a thin disk.

 There are, however, subtle differences between 
the kinematics of these two galaxies and a typical elliptical galaxy. First, 
in the two galaxies studied here, the values of both the rotation velocity and 
the  velocity dispersion are high, but the dispersion  is generally larger 
($V_{\mathrm c} \leq \sigma$) - compare Figs. 1b and 1c,
and Figs. 2b and 2c. Thus, the support against 
gravitational collapse due to both pressure and rotation is important though 
the pressure-support dominates  over a large 
radial range. 
In a typical bright giant elliptical, on the other hand, the values of 
rotation velocity seen are much smaller  $\sim$ few $\times$ 10 km s$^{-1}$ 
(Binney \& Merrifield 1998), thus the support 
due to rotation is negligible. The kinematics of the two
galaxies studied here is also 
in a stark contrast to a typical spiral disk which is supported by rotation.

Second, the velocity dispersion increases monotonically with
radius in the outer parts in both Arp 224 and Arp 214, unlike in
a typical elliptical galaxy (Binney \& Merrifield 1998). This is
probably a sign of its tidal origin and it also points to the
existence of an outer, perhaps unrelaxed component which shows the
exponential mass profile. We obtained the ellipticity of the
isophotal contours from the $K_{\mathrm s}$ band images  for these galaxies. 
Interestingly, we find that in both cases, the ellipticity shows a sharp 
discontinuity at the region where
the velocity dispersion begins to rise again, at $\sim 10''$ (on the 
positive side of the major axis) for Arp 224 and at $\sim  20''$ (on both the 
 sides) for Arp 214. This is also the region beyond which
the exponential disk gives a good fit (see Figs. 1a and 2a). 
 The ellipticity drops from a maximum of 0.18 at $14''$ to 0.07 at $17''$
in Arp 224, while  it drops from a maximum
of 0.14 at $19''$ to 0.05 at $21''$ in Arp 214. This gives an independent 
confirmation of a separate structural component (a disk) in the outer parts
of the galaxy, outside of the central bulge.
Note that this is approximately the radius at which the rotation curve shows 
a maximum.
The coincidence of the rise in the velocity dispersion and the
fall in the rotation curve in both these cases can be understood physically
 as arising due to the phenomenon of
asymmetric drift (Binney \& Tremaine 1987). This aspect will be studied 
in a future paper.

We note that in detail the kinematics show a complex
behaviour. For example, the rotation curve is  asymmetric 
on the two sides of the major axis for Arp 224 (Fig. 1b). This is probably 
due to the strong disturbance that the galaxy has undergone, and
is not surprising. Note that even normal galaxies show a significant 
rotational asymmetry of a lopsided nature, 
representing the effects of a past tidal encounter (Jog 2002).

In Arp 214, the HI gas shows disturbed kinematics believed
to be due to a projection of the outer warped and twisted  disk
onto the inner regions (Schwarz 1985). This model can be ruled
out for the stellar kinematics since the latter spatially  cover a
central region even smaller than the beam-size of the HI study.
Also, the gas and stars are often decoupled in mergers (Genzel
et al. 2001), mainly because of the dissipational nature of gas.
In Arp 224, any incoming tails (Hibbard \& Mihos 1995), if
aligned along the line-of-sight, could give rise to a spurious dispersion
but it would need a contrived geometry for this scenario to
explain the details of the dispersion profile observed. Thus, both these 
alternatives for the origin of the high velocity dispersion ($\sigma$) 
observed can be ruled out.

Further, the symmetric
central profile of $\sigma$ (as in an elliptical), the increase
in $\sigma$ with radius in the outer parts where the exponential
disk fits well, and the correlation
between the falling rotation curve and the increasing $\sigma$,
convincingly show that the values of $\sigma$ observed in Figs.
1c and 2c represent true stellar random motion, and are an
intrinsic property of the galaxy. Thus, our interpretation of
these systems as being mainly pressure-supported is well-justified.

So far for simplicity we have portrayed spiral and elliptical
galaxies as being supported respectively by rotation and random
motion alone. In reality both types can sometimes show a range
of dynamical and photometric properties. For example, the
low-luminosity ellipticals are often disky (Faber et al. 1997),
and show a significant rotational support with $V_{\mathrm c} / \sigma$ 
between 0.5-1.0 (Bender et al. 1992), and can show
exponential luminosity profiles (Caon et al. 1993). 
On the other hand, bulges in isolated spiral galaxies are
dynamically hot (Binney \& Merrifield 1998), and sometimes show
a comparable support from both rotation and random motion
(Bender et al. 1992). This behaviour is not understood, and the present 
study on mergers may help shed some light on this.

In order to better understand the mass distribution and evolution
of the thirteen mergers showing exponential profiles (Chitre \& Jog 2002),
we compare their photometric properties with those of
typical isolated spirals.
We find that the  distribution of the $K_{\mathrm s}$ band disk
scalelengths of these is similar to that for a sample of undisturbed
spiral galaxies (Peletier et al. 1994). Such a similarity was
noted earlier (Schwarzkopf \& Dettmar 2000), but only for a
merger of galaxies with a mass ratio less than 1:10.
Next, a simultaneous disk plus an $r^{1/4}$ profile for the bulge 
was tried, but this overestimates the
luminosity in the middle radial range, and hence is not a good
fit for any of the thirteen galaxies.
A Sersic or a generalized de Vaucouleurs fit with an $r^{1/n}$ profile 
(e.g., Caon et al. 1993) was 
also attempted, but the resulting value of $n$ is extremely sensitive to the 
radial range chosen, hence this fit cannot be used.
The disk scalelengths for our sample  galaxies  are
comparable to those for isolated spirals as shown above, hence
these are too large compared to those of typical S0 galaxies
(Binney \& Tremaine 1987). 

It is interesting that the disky ellipticals
show (Scorza \& Bender 1995) smaller scalelengths than S0's or Sa's
  while our sample shows values similar to normal spirals,
implying different dynamical evolution of the disky ellipticals
compared to our sample. The idea of a different origin is also
supported by the dynamical data presented here, because Arp 224 and Arp 214
are mainly pressure-supported whereas the low-luminosity, disky
ellipticals studied by Rix et al (1999) show a dominant rotational 
support beyond one effective radius.
Another difference is that the Rix et al. (1999) sample  
galaxies are less luminous than $M_{\mathrm B}$ = - 19.5, whereas  
our galaxies are brighter than this. 

\bigskip

\noindent {\bf 3. Dynamical Implications and Puzzles} 

The results in Sect. 2 show that despite the evidence for a strong tidal 
interaction such as the high velocity dispersion, tidal tails,
and a disturbed appearance as shown by these galaxies, 
 their radial mass distribution seems largely unaffected.
However, because of the high velocity dispersion, the disk is
probably puffed up vertically, and also would not show large-scale
spiral structure (Binney \& Tremaine 1987). Our work shows that the
galaxy disks are not fragile as is sometimes suggested from
their vertical heating during an encounter (Toth \&
Ostriker 1992), in fact if anything 
the galaxies are robust in their radial mass distribution.

 These mergers seem to have  avoided suffering a complete violent relaxation 
(Barnes 1992) that would have resulted in an $r^{1/4}$ de Vaucouleurs type of 
profile (de Vaucouleurs 1977) as seen in an
elliptical galaxy.
Yet these galaxies are indistinguishable (Chitre \& Jog 2002) in appearance 
from the
mergers which show an $r^{1/4}$ profile. Thus the origin of these
galaxies and their dynamics is a puzzle. The standard indicators
of a galaxy type, namely the radial mass distribution and the kinematics,
do not tally in these galaxies.  This may indicate that these galaxies 
are in transition from a spiral galaxy to an early-type spheroidal system, 
and hence have properties interim to both of these types.

We suggest that future observations that give kinematic data for the other 
eleven mergers with the exponential profiles would be extremely useful. This 
will help establish this new class of mergers on a firmer footing.

How do these mergers compare with the theoretical models? 
The theoretical work involving N-body simulations of mergers of
galaxies has a vast range of parameter space available covering
different progenitor mass ratios, encounter geometry, and
inclination of galaxies, all of which has not yet been fully
sampled.  We find that there are no analogs of such 
pressure-supported  remnants with exponential
profiles in the models in the literature so far.
Mergers of galaxies with equal mass (Barnes 1992), and comparable
masses  with a mass ratio in the range of 1:4-1:1 (Naab \& Burkert 2001, 
Bendo \& Barnes 2000), have been studied.
These result in a pressure-supported remnant, with an $r^{1/4}$ profile, as 
seen in a bright giant elliptical galaxy- and different from our sample 
(Sect. 2). These models were largely motivated by the observations of
ultraluminous galaxies, which involve mergers of comparable-mass
galaxies. However, the models with the mass ratios
1:4 and 1:3 cannot yet be excluded, because
for some input parameters these produce remnants with
properties overlapping with those observed for Arp 224 and Arp 214. First,
these remnants can have comparable support from rotation and
random motion, with $V_{\mathrm c} / \sigma \sim $ 0.5-1  at
one effective radius (Naab et al. 1999, Naab \&
Burkert 2001, Cretton et al. 2001, Barnes 1998). This range is roughly
comparable to the observed data presented in our work. Second, the
unpublished surface density results for the models of Naab \&
Burkert (Naab 2002, personal communication) show a trend to more disk-like 
profiles in the outer regions.

At the other end of the mass ratio, the studies of a satellite merging with a 
parent galaxy with a mass ratio of about 1:10 (Quinn et al.
1993, Walker et al. 1996, Velazquez \& White 1999) show 
an increase in the random velocity at large radii along all the three 
directions in a galaxy, but their values obtained are 
smaller by a factor of 2-3 compared to the velocity dispersion values observed 
already at a galactic radius of a few kpc in Arp 224 and Arp 214.

We suggest that future N-body work should explore the new parameter
range covering mergers of a massive
satellite with the parent galaxy with values of mass-ratios not
covered so far, namely between 1:10-1:4. The simulations
of mergers with this mass range are currently lacking. This range
 can result in a higher increase in velocity dispersion
than in the satellite accretion work studied so far, yet avoid
the full-scale violent relaxation seen in mergers of
comparable-mass galaxies so that
the exponential distribution is unaffected. Physically, 
the above idea has the potential of explaining the
unusual, mixed set of properties shown by Arp 224 and Arp 214.
An inclusion of gas with the associated dissipation may also be 
important, but it should have a strong enough effect to affect the 
distribution of the main mass component namely the old stars that 
we have studied via the near-infrared.

\bigskip

\noindent {\bf 4. Summary and Discussion} 

High random velocities in mergers are expected
theoretically (e.g., Walker et al. 1996), however,
it is the combination of exponential profiles and the dominance
of velocity dispersion over rotation that we have found here
that is unexpected and hard to explain. Thus, the photometric near-IR 
properties for a new type of sample chosen mainly from its 
morphologically disturbed appearance (Chitre \& Jog 2002), combined with the 
kinematic data studied in this paper, have together led to a complete and 
more intriguing picture of these mergers than either data alone
would have.
The origin and evolution of this new class of observed
merger remnants is an open question, and needs further observational and
dynamical studies.

The new mass range of 1:10-1:4 proposed here is likely to be
common in mergers at high redshift, as shown in the hierarchical
merging models (Steinmetz \& Navarro 2002) for galaxy formation.
Observationally, it is
well-known that galaxy morphology evolves with redshift (Abraham
\& van den Bergh 2001). The two merger remnants studied here appear to
be the present-day analogs of the high percentage of peculiar
galaxies without well-developed spiral structure observed to be
common at high redshift. Thus, the present study may be relevant for
the early evolution of galaxies.

\bigskip

\noindent {\bf {Acknowledgements}}

This publication makes use of archival data products from the Two Micron
All Sky Survey (http://isra.ipac.caltech.edu), and the HYPERCAT archival
database (http://www-obs.univ-lyon1.fr/hypercat/). 

We would like to thank the referee, T. Naab, for 
constructive comments; and C.A. Narayan for useful discussions.

\bigskip

\bigskip

\noindent{\bf References}

\bigskip

\noindent Abraham, R.G., \& van den Bergh, S. 2001, Science, 293, 1273

\noindent Barnes, J.E. 1992, ApJ, 393, 484

\noindent Barnes, J.E. 1998, in Galaxies: Interactions and
Induced Star Fromation, Saas-Fee Advanced course 26, ed. D.
Friedli, L. Martinet, \& D. Pfenniger (Berlin: Springer), 275

\noindent Bender, R., Burstein, D., \& Faber, S.M. 1992, ApJ, 399, 462

\noindent Bendo, G.J., \& Barnes, J.E. 2000, MNRAS, 316, 315

\noindent Binney, J., \& Merrifield, M. 1998, Galactic Astronomy
(Princeton: Princeton Univ. Press)

\noindent Binney, J., \& Tremaine, S. 1987, Galactic Dynamics
(Princeton: Princeton Univ. Press)

\noindent Caon, N., Capaccioli, M., \& D'Onofrio, M. 1993,
MNRAS, 265, 1013

\noindent Chitre, A., \& Jog, C.J. 2002, A\&A, 388, 407

\noindent Cretton, N., Naab, T., Rix, H.-W., \& Burkert, A. 2001, ApJ, 554, 291

\noindent de Vaucouleurs, G. 1977, in  Evolution of Galaxies and
Stellar Populations, ed. B.M. Tinsley, \& R.B. Larson (New Haven: 
Yale Univ. Obs.), 43

\noindent Faber, S.M.  et al. 1997, AJ, 114, 1771

\noindent Freeman, K.C. 1970, ApJ, 160, 811

\noindent Genzel, R., Tacconi, L.J., Rigopoulou, D., Lutz, D., \&
Tecza, M.   2001, ApJ, 563, 527

\noindent Heraudeau, Ph., \& Simien, F. 1998, A\&AS, 133, 317

\noindent Hibbard, J. E., \& Mihos, J.C. 1995, AJ, 110, 140

\noindent Jog, C.J. 2002, A\&A, 391, 471

\noindent Naab, T., \& Burkert, A. 2001, in Galaxy Disks and Disk Galaxies,  
ASP Conf. Ser. 230,  ed. J.G. Funes \& E.M. Corsini
(San Francisco: ASP), 453

\noindent Naab, T., Burkert, A., \& Hernquist, L.  1999, ApJ, 523, L133

\noindent Peletier, R.F., Valentijn, E.A., Moorwood, A.F.M., \&
Freudling, W.   1994, A\&AS, 108, 621

\noindent Quinn, P.J., Hernquist, L., \& Fullagar, D.P. 1993,
ApJ, 403, 74

\noindent Rix, H.-W., Carollo, C.M., \& Freeman, K.C. 1999, ApJ,
513, L25

\noindent Schwarz, U.J. 1985, A\&A, 142, 273

\noindent Schwarzkopf, U., \& Dettmar, R.-J. 2000, A\&A, 361, 451

\noindent Schweizer, F. 1982, ApJ, 252, 455

\noindent Scorza, C., \& Bender, R. 1995, A\&A, 293, 20

\noindent Scoville, N.Z. et al.  2000, AJ, 119, 991

\noindent Simien, F., \& Prugniel, Ph. 1997, A\&AS, 126, 15

\noindent Stanford, S.A., \& Bushouse, H.A., 1991, ApJ, 371, 92

\noindent Steinmetz, M., \& Navarro, J.F. 2002, New A, 7, 155

\noindent Toth, G., \& Ostriker, J.P. 1992, ApJ, 389, 5

\noindent Velazquez, H., \& White, S.D.M. 1999, MNRAS, 304, 254

\noindent Walker, I.R., Mihos, J.C., \& Hernquist, L. 1996, ApJ, 460, 121

\noindent Wright, G.S., James, P.A., Joseph, R.D., \& McLean,
J.S.   1990, Nature, 344, 417

\newpage

\centerline {\bf Figure Captions}

\noindent {\bf Figure 1.} Photometric (a) and kinematic (b and c) 
 properties of Arp 224. (a) shows the surface brightness
distribution with radius in the $K_{\mathrm s}$ band derived using the image
from the 2MASS data. The inverted triangles denote the observed
values and the straight line is obtained by fitting an
exponential to the outer points. The exponential disk gives a
good fit beyond $10''$ (where $1''$ = 0.38 kpc). The disk scalelength
is $7.6''$ or  2.91 kpc. (b) shows $V_{\mathrm c}$, the circular rotation 
velocity of Arp 224 on either side of the galaxy centre. The rotation
velocity reaches a maximum value at around $10''$ and then
decreases beyond this radius. (c) shows $\sigma$, the velocity
dispersion. It peaks at the centre with a value of $\sim$ 210 
km s$^{-1}$ and then dips gradually upto $\sim $ $10''$ on either side, 
and again increases beyond this radius on the positive side of the axis.
 $V_{\mathrm c}$ is less than $\sigma$ at all radii covered by the 
 kinematical data.

\noindent {\bf Figure 2.} Photometric (a) and kinematic (b and c) properties 
of Arp 214, with details similar to Fig. 1.
(a) The exponential disk gives a
good fit beyond $20''$ (where $1''$ = 0.065 kpc). The disk scalelength
is $26.5''$ or 1.72 kpc. (b) $V_{\mathrm c}$, the rotation
velocity reaches a maximum value at $20''$ and then
decreases beyond this radius. (c) $\sigma$, the velocity
dispersion peaks at the centre with a value of $\sim$ 180 km s$^{-1}$ and 
falls on either side, and again
increases beyond about $20''$. Between $10''- 20''$, 
 $V_{\mathrm c}$  is larger than
$\sigma$, beyond $20''$ the dispersion dominates over rotation again.

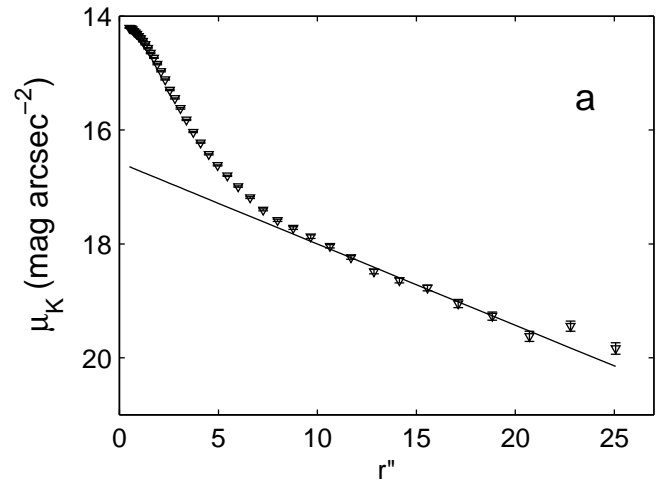
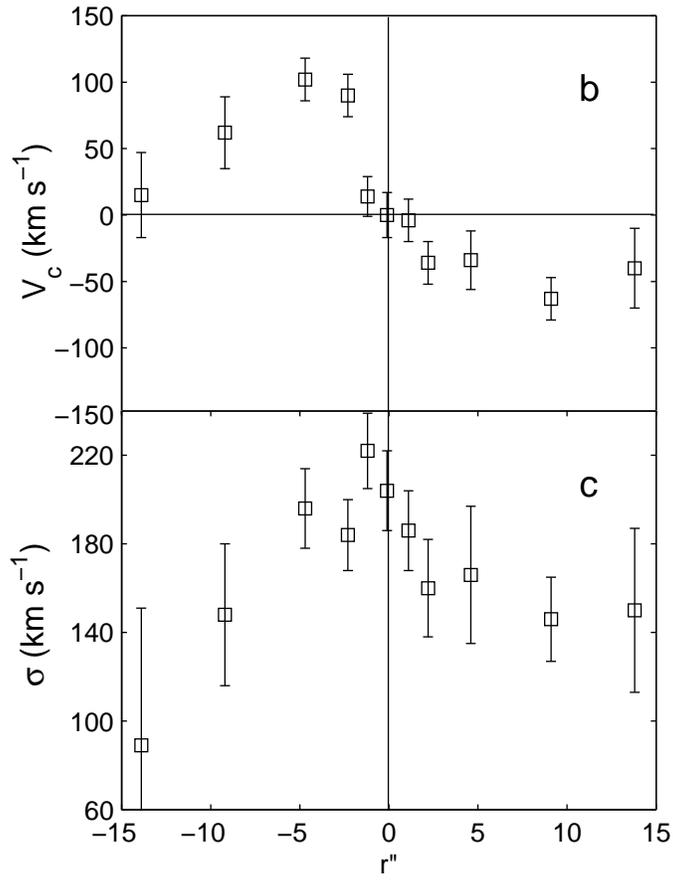
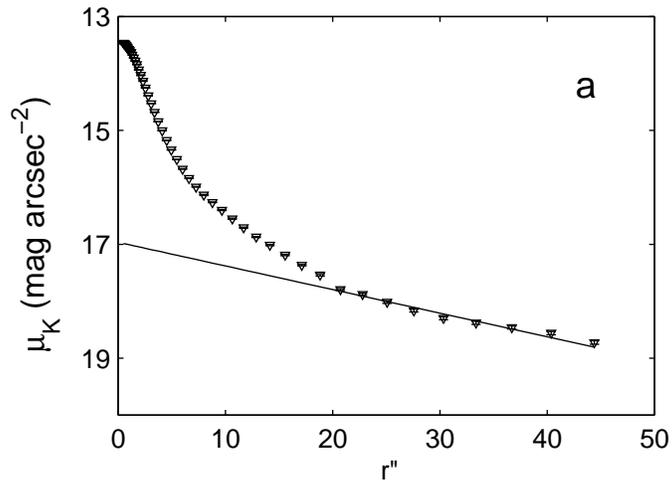
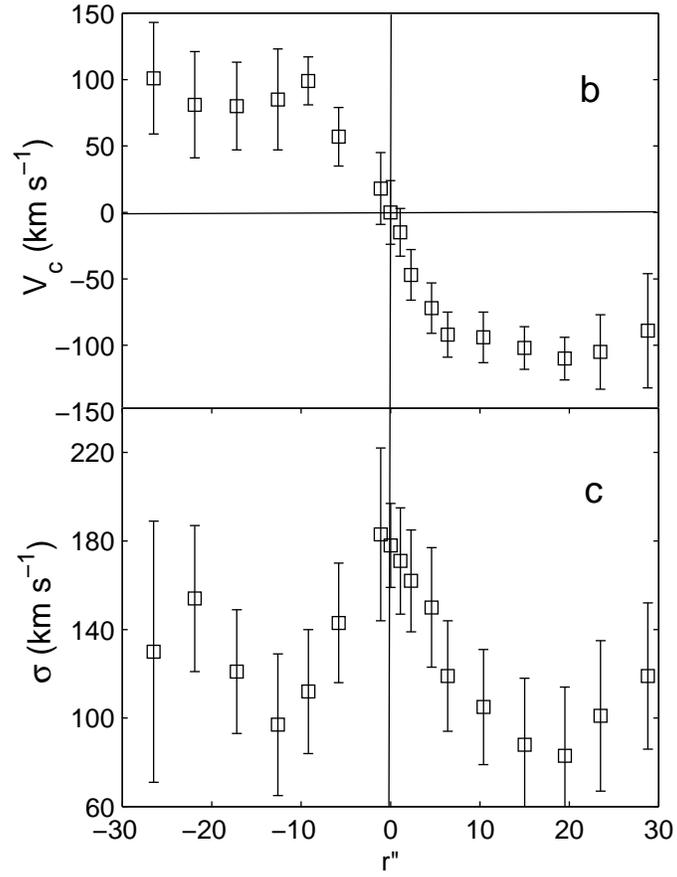
\end{document}